\documentclass[12pt]{article}
\usepackage{ametsoc}
\usepackage{amsmath} 
\usepackage{amssymb} 
\usepackage{amsfonts}
\usepackage{graphicx,color}                 
\usepackage{subfigure}
\usepackage{lineno}
\usepackage{xcolor}
\usepackage{multirow}
\linenumbers

\newcommand{\myabstract}{

}
\begin{document}
%
%
\title{\textbf{\large{Covariance inflation in the ensemble Kalman filter: a residual nudging perspective and some implications}}}
%
%
\author{\textsc{Xiaodong Luo}
				\thanks{\textit{Corresponding author address:}
				International Research Institute Of Stavanger (IRIS), Thorm{\o}hlens Gate 55, 5008 Bergen, Norway
				\newline{E-mail: xiaodong.luo@iris.no}} \\
				\textit{\footnotesize{International Research Institute Of Stavanger (IRIS), 5008 Bergen, Norway}}				
\and
\centerline{\textsc{Ibrahim Hoteit}}\\
\centerline{\textit{\footnotesize{King Abdullah University of Science and Technology (KAUST), Thuwal 23955-6900, Saudi Arabia}}}
}
%

\ifthenelse{\boolean{dc}}
{
\twocolumn[
\begin{@twocolumnfalse}
\amstitle

\begin{center}
\begin{minipage}{13.0cm}
\begin{abstract}
	\myabstract
	\newline
	\begin{center}
		\rule{38mm}{0.2mm}
	\end{center}
\end{abstract}
\end{minipage}
\end{center}
\end{@twocolumnfalse}
]
}
{
\amstitle
\begin{abstract}
\myabstract {\color{black}{ This note examines the influence of covariance inflation on the distance between the measured observation and the simulated (or predicted) observation with respect to the state estimate. In order for the aforementioned distance to be bounded in a certain interval, some sufficient conditions are derived, indicating that the covariance inflation factor should be bounded in a certain interval, and that the inflation bounds are related to the maximum and minimum eigenvalues of certain matrices. Implications of these analytic results are discussed, and a numerical experiment is presented to verify the validity of our analysis.} }
\end{abstract}
}

\section{Data assimilation with residual nudging}
A finite, often small, ensemble size has some well known effects that may substantially influence the behaviour of an ensemble Kalman filter (EnKF). These effects include, for instance, {\color{black}{rank deficient sample error covariance matrices, systematically underestimated error variances, and in contrast, exceedingly large error cross-covariances of the model state variables \citep{Whitaker-ensemble}}}. In the literature, the latter two issues are often tackled through covariance localization \citep{Hamill-distance}, while the first issue, under-estimation of sample {\color{black}{variances}}, is often handled by covariance inflation \citep{Anderson-Monte}, in which one artificially increases the sample {\color{black}{variances}}, either multiplicatively (see, for example, \citealp{Anderson-Monte,Anderson2007,Anderson2009,Bocquet2012-combining,Miyoshi2011-Gaussian}), or additively (see, for example, \citealp{Hamill2011-what}), or in a hybrid way by combining both multiplicative and additive inflation methods (see, for example, \citealp{Whitaker2012-evaluating}), or through other ways such as relaxation to the prior \citep{zhang2004impacts}, multi-scheme ensembles \citep{meng2007tests}, modification of the eigenvalues of sample error covariance matrices \citep{Altaf2013-improving,Luo2011_EnLHF,Ott-local,Triantafyllou2012-assessing}, back projection of the residuals to construct new ensemble members \cite{Song2010-adaptive} to name but a few. In general, covariance inflation tends to increase the robustness of the EnKF against uncertainties in data assimilation \citep{Luo2011_EnLHF}, and often also improves the filter performance in terms of estimation accuracy.

The focus of this note is to study the effect of covariance inflation from the point of view of residual nudging \citep{Luo2012-residual}. Here, the ``residual'' with respect to an $m$-dimensional system state $\mathbf{x}$ is a vector in the observation space, defined as $\mathbf{H} \mathbf{x} - \mathbf{y}$ \footnote{\color{black}{In the literature, the vector with the opposite sign, $\mathbf{y} -\mathbf{H} \mathbf{x}$, is often called ``innovation''.}}, where $\mathbf{H}: \mathbb{R}^m \rightarrow \mathbb{R}^p$ is a linear observation operator, and $\mathbf{y}$ the corresponding $p$-dimensional observation vector. Throughout this note, our discussion is confined to the filtering (or analysis) step of the EnKF, so that the time index in the EnKF is dropped. The linearity assumption in the observation operator $\mathbf{H}$ is taken in order to simplify our discussion. The result to be presented later, though, might also provide insights into more complex situations.

Before introducing the concept of residual nudging, let us define some additional notations. We assume that the observation system is given by
\begin{linenomath*}
\begin{equation} \label{eq:obs_system}
\mathbf{y} = \mathbf{H} \mathbf{x} + \mathbf{v} \, ,
\end{equation}
\end{linenomath*}
where $\mathbf{v}$ is the vector of observation error, with zero mean and a non-singular covariance matrix $\mathbf{R}$. We further decompose $\mathbf{R}$ as $\mathbf{R} = \mathbf{R}^{1/2} \, \mathbf{R}^{T/2}$, where $\mathbf{R}^{1/2}$ is a non-singular square root of $\mathbf{R}$ and $\mathbf{R}^{T/2}$ denotes the transpose of $\mathbf{R}^{1/2}$.

To measure the length of a vector $\mathbf{z}$ in the observation space, we adopt the following weighted Euclidean norm
\begin{linenomath*}
\begin{equation} \label{eq:l2_norm}
\Vert \mathbf{z} \Vert_{\mathbf{R}} \equiv \sqrt{\mathbf{z}^T \, \mathbf{R}^{-1} \, \mathbf{z}} \, .
\end{equation}
\end{linenomath*}
One may convert the weighted Euclidean norm to the standard Euclidean norm by noticing that $\Vert \mathbf{z} \Vert_{\mathbf{R}} = \Vert \mathbf{R}^{-1/2} \, \mathbf{z} \Vert_{2}$, where $\Vert \bullet \Vert_{2}$ denotes the standard Euclidean norm. As a result, many topological properties with respect to the standard Euclidean norm, e.g., the triangle inequality (see (\ref{eq:residual_norm_triangle_ineq}) below), still hold with respect to the weighted Euclidean norm.

The idea of data assimilation with residual nudging (DARN) is the following. Let $\mathbf{x}^{tr}$ be the true system state (truth), $\mathbf{y}^{o} = \mathbf{H} \mathbf{x}^{tr} + \mathbf{v}^o$ the recorded observation for a specific realization $\mathbf{v}^o$ of the observation error, and $\hat{\mathbf{x}}$ the state estimate (e.g., either the prior or posterior estimate) obtained from a data assimilation (DA) algorithm. Then the residual $\hat{\mathbf{r}} = \mathbf{H} \hat{\mathbf{x}} - \mathbf{y}^{o} =  \mathbf{H} \hat{\mathbf{x}} - \mathbf{H} \mathbf{x}^{tr} - \mathbf{v}^o$. By the triangle inequality, the weighted Euclidean norm of the residual (residual norm hereafter) satisfies
\begin{linenomath*}
\begin{equation} \label{eq:residual_norm_triangle_ineq}
\Vert \hat{\mathbf{r}} \Vert_{\mathbf{R}} \leq \Vert \mathbf{H} \hat{\mathbf{x}} - \mathbf{H} \mathbf{x}^{tr} \Vert_{\mathbf{R}} + \Vert \mathbf{v}^o \Vert_{\mathbf{R}}\, .
\end{equation}
\end{linenomath*}
If the DA algorithm performs reasonably well, one may expect that the magnitude of $\Vert \mathbf{H} \hat{\mathbf{x}} - \mathbf{H} \mathbf{x}^{tr} \Vert_{\mathbf{R}}$ not be significantly larger than $\Vert \mathbf{v}^o \Vert_{\mathbf{R}}$. As a result, one may obtain an upper bound of $\Vert \hat{\mathbf{r}} \Vert_{\mathbf{R}}$ in terms of $\Vert \mathbf{v}^o \Vert_{\mathbf{R}}$, e.g, in the form of $\beta \Vert \mathbf{v}^o \Vert_{\mathbf{R}}$, where $\beta$ is a non-negative scalar coefficient. In practice, though, $\Vert \mathbf{v}^o \Vert_{\mathbf{R}}$ is often unknown. As a remedy, we replace $\Vert \mathbf{v}^o \Vert_{\mathbf{R}}$ by an upper bound of the expectation $\mathbb{E} (\Vert \mathbf{v} \Vert_{\mathbf{R}})$ of the weighted Euclidean norm of the observation error $\mathbf{v}$, where $\mathbb{E}$ denotes the expectation operator. One such upper bound can be obtained by noticing that
\begin{linenomath*}
\begin{equation} \label{eq:expectation_ineq}
\left( \mathbb{E} (\Vert \mathbf{v} \Vert_{\mathbf{R}}) \right)^2 \leq \mathbb{E} (\Vert \mathbf{v} \Vert_{\mathbf{R}}^2) = \text{trace} \left(\mathbf{R}^{-1} \mathbb{E} (\mathbf{v} \mathbf{v}^T) \right) = \text{trace}(\mathbf{I}_p) = p \, ,
\end{equation}
\end{linenomath*}
where the operator ``$\text{trace}$'' evaluates the trace of a matrix, and $\mathbf{I}_p$ the $p$-dimensional identity matrix. From (\ref{eq:expectation_ineq}), we have the upper bound $ \mathbb{E} (\Vert \mathbf{v} \Vert_{\mathbf{R}}) \leq \sqrt{p}$. Consequently, we want to find a state estimate $\hat{\mathbf{x}}$ whose residual norm $\Vert \hat{\mathbf{r}} \Vert_{\mathbf{R}}$ satisfies
\begin{linenomath*}
\begin{equation} \label{eq:residual_norm_target_ineq}
\Vert \hat{\mathbf{r}} \Vert_{\mathbf{R}} \leq  \beta \sqrt{p}
\end{equation}
\end{linenomath*}
for a pre-chosen $\beta$. It is worthy of mentioning that in general it may be difficult to identity which $\beta$ gives the best state estimation accuracy with respect to the truth $\mathbf{x}^{tr}$. Therefore, in \citet{Luo2012-residual} we mainly used DARN as a safeguard strategy, that is, if a state estimate $\hat{\mathbf{x}}$ is found to have a too large residual norm, then we try to introduce some correction to the state estimate in order to reduce its residual norm, which in turn might also improve the estimation accuracy.

In \citet{Luo2012-residual} we introduced DARN to the analysis $\hat{\mathbf{x}}^a$ in the ensemble adjustment Kalman filter (EAKF, see \citealp{Anderson-ensemble}). In the EAKF with residual nudging (EAKF-RN), if the residual norm of $\hat{\mathbf{x}}^a$ is less than $\beta \sqrt{p}$, then we accept $\hat{\mathbf{x}}^a$ as a reasonable estimate and no change is made. Otherwise, a correction is introduced to $\hat{\mathbf{x}}^a$ in a way such that the residual norm of the modified state estimate $\tilde{\mathbf{x}}^a$ is exactly $\beta \sqrt{p}$, and that among all possible state estimates whose residual norms are equal to $\beta \sqrt{p}$, the simulated (or predicted) observation $\mathbf{H} \tilde{\mathbf{x}}^a$ of the modified state estimate $\tilde{\mathbf{x}}^a$ has the shortest distance to the one $\mathbf{H} \hat{\mathbf{x}}^a$ of the original state estimate $\hat{\mathbf{x}}^a$. Numerical results in \citet{Luo2012-residual} show that the EAKF-RN exhibits (sometimes substantially) improved filter performance, in terms of estimation accuracy and/or stability against filter divergence, compared to the EAKF.
Extension of DARN to other types of filters is also possible, for example, see \citet{Luo2013-efficient}.

\section{Covariance inflation from the point of view of residual nudging} \label{sec:inflation}
Here we examine the effect of covariance inflation on the analysis residual norm. To this end, we first recall that the mean update formula in the EnKF (without perturbing the observation) is given by
\begin{linenomath*}
\begin{equation} \label{eq:mean_update}
\begin{split}
& \hat{\mathbf{x}}^a = \hat{\mathbf{x}}^b + \mathbf{K} \left(\mathbf{y}^o - \mathbf{H} \hat{\mathbf{x}}^b  \right) \, , \\
& \mathbf{K} = \hat{\mathbf{C}}^b \mathbf{H}^T \left( \mathbf{H} \hat{\mathbf{C}}^b \mathbf{H}^T + \mathbf{R} \right)^{-1} \, ,
\end{split}
\end{equation}
\end{linenomath*}
where $\hat{\mathbf{x}}^b$ and $\hat{\mathbf{x}}^a$ are the sample means of the background and analysis ensembles, respectively; $\mathbf{K}$ is the Kalman gain; and $\hat{\mathbf{C}}^b$ is a certain symmetric, positive semi-definite matrix in accordance to the chosen inflation scheme. In general $\hat{\mathbf{C}}^b$ may be related, but not necessarily proportional, to the sample error covariance matrix $\hat{\mathbf{P}}^b$ of the background ensemble. For instance, in the hybrid EnKF $\hat{\mathbf{C}}^b$ can be a mixture of $\hat{\mathbf{P}}^b$ and a ``background covariance'' $\mathbf{B}$ \citep{Hamill-hybrid}, or partially time-varying as in \citet{Hoteit2002}.

Our objective is to examine under which conditions the residual norm $\Vert \hat{\mathbf{r}}^a \Vert_{\mathbf{R}}$ of the analysis $\hat{\mathbf{x}}^a$ satisfies $ \beta_l \, \sqrt{p} \leq \Vert \hat{\mathbf{r}}^a \Vert_{\mathbf{R}} \leq \beta_u \, \sqrt{p}$, where $\beta_l$ and $\beta_u$ ($0 \leq \beta_l \leq \beta_u$) represents the lower and upper values of $\beta$ that one wants to set for the analysis residual norm in DARN. Different from the previous works \citep{Luo2012-residual,Luo2013-efficient}, the lower bound $\beta_l \, \sqrt{p}$ is introduced here in order to make our discussion below slightly more general. In practice it may also be used to prevent too small residual norms in certain circumstances in order to avoid, for instance, a state estimate that over-fits the observation, a phenomenon that may be caused by ``over-inflation'', as will be shown later.

Inserting Eq. (\ref{eq:mean_update}) into $\hat{\mathbf{r}}^a = \mathbf{H} \hat{\mathbf{x}}^a - \mathbf{y}^o$, one has
\begin{linenomath*}
\begin{equation} \label{eq:residual_update}
\hat{\mathbf{r}}^a =  \mathbf{R} \left(\mathbf{H} \hat{\mathbf{C}}^b \mathbf{H}^T + \mathbf{R} \right)^{-1} \hat{\mathbf{r}}^b \, ,
\end{equation}
\end{linenomath*}
where $\hat{\mathbf{r}}^b = \mathbf{H} \hat{\mathbf{x}}^b - \mathbf{y}^o$. Multiplying both sides of Eq. (\ref{eq:residual_update}) by $\mathbf{R}^{-1/2}$, one obtains
\begin{linenomath*}
\begin{equation} \label{eq:weighted_residual_update}
(\mathbf{R}^{-1/2} \hat{\mathbf{r}}^a) = \left( \mathbf{R}^{-1/2}\mathbf{H} \hat{\mathbf{C}}^b \mathbf{H}^T \mathbf{R}^{-T/2} + \mathbf{I}_p \right)^{-1} (\mathbf{R}^{-1/2} \hat{\mathbf{r}}^b ) \, .
\end{equation}
\end{linenomath*}
To derive the bounded residual norm, we first consider under which conditions the upper bound $\Vert \hat{\mathbf{r}}^a \Vert_{\mathbf{R}} \leq \beta_u \, \sqrt{p}$ is guaranteed to hold. Given that (cf (\ref{eq:mtx_ineq_1}) later)
\begin{linenomath*}
\begin{equation} \label{eq:residual_normal_update}
\Vert \hat{\mathbf{r}}^a \Vert_{\mathbf{R}} = \Vert \mathbf{R}^{-1/2} \hat{\mathbf{r}}^a \Vert_2 \leq \Vert (\mathbf{R}^{-1/2}\mathbf{H} \hat{\mathbf{C}}^b \mathbf{H}^T \mathbf{R}^{-T/2} + \mathbf{I}_p )^{-1} \Vert_2 \; \Vert \hat{\mathbf{r}}^b \Vert_{\mathbf{R}} \, ,
\end{equation}
\end{linenomath*}
a sufficient condition is thus
\begin{linenomath*}
\begin{equation} \label{eq:upper_bound_sufficient_cond_no_inf}
\Vert (\mathbf{R}^{-1/2}\mathbf{H} \hat{\mathbf{C}}^b \mathbf{H}^T \mathbf{R}^{-T/2} + \mathbf{I}_p )^{-1} \Vert_2 \leq  \dfrac{\beta_u \, \sqrt{p}}{\Vert \hat{\mathbf{r}}^b \Vert_{\mathbf{R}}} \, .
\end{equation}
\end{linenomath*}

Let
\begin{linenomath*}
\begin{equation} \label{eq:mtx_A}
\mathbf{A} = \mathbf{R}^{-1/2}\mathbf{H} \hat{\mathbf{C}}^b \mathbf{H}^T \mathbf{R}^{-T/2} \, ,
\end{equation}
\end{linenomath*}
and $\lambda_{max}$ and $\lambda_{min}$ be the maximum and minimum eigenvalues of $\mathbf{A}$, respectively. Recalling that the induced 2-norm of a symmetric positive semi-definite matrix is exactly the maximum eigenvalue of that matrix \citep[\S5.6.6]{Horn1990-matrix}, we have
\begin{linenomath*}
\begin{equation} \label{eq:2_norm_no_inf}
\Vert (\mathbf{A} + \mathbf{I}_p )^{-1} \Vert_2 = (\lambda_{min} + 1)^{-1} \, .
\end{equation}
\end{linenomath*}
Therefore (\ref{eq:upper_bound_sufficient_cond_no_inf}) leads to
\begin{linenomath*}
\begin{equation} \label{eq:lambda_min_no_inf}
\lambda_{min} + 1 \geq \dfrac{ \Vert \hat{\mathbf{r}}^b \Vert_{\mathbf{R}}}{\beta_u \, \sqrt{p}} \, .
\end{equation}
\end{linenomath*}
If $\Vert \hat{\mathbf{r}}^b \Vert_{\mathbf{R}}$ is relatively small such that $\Vert \hat{\mathbf{r}}^b \Vert_{\mathbf{R}} \leq \beta_u \, \sqrt{p}$, then (\ref{eq:lambda_min_no_inf}) automatically holds. However, if $\Vert \hat{\mathbf{r}}^b \Vert_{\mathbf{R}} > \beta_u \, \sqrt{p}$, and that $\lambda_{min}$ is very small, then there is no guarantee that (\ref{eq:lambda_min_no_inf}) will hold. A small $\lambda_{min}$ may appear, for instance, when the ensemble size $n$ is smaller than the dimension $p$ of the observation space. In such circumstances, the matrix $\mathbf{A}$ may be singular with $\lambda_{min} = 0$, and the singularity may not be avoided only through the multiplicative covariance inflation. If one cannot afford to increase the ensemble size $n$, then a few alternative strategies may be adopted to address (or at least mitigate) the problem of singularity. These include, for instance, (a) introducing covariance localization \citep{Hamill-distance} to $\hat{\mathbf{P}}^b$ in order to increase its rank \citep{Hamill2009}; (b) replacing the sample error covariance $\hat{\mathbf{P}}^b$ by a hybrid of $\hat{\mathbf{P}}^b$ and some full-rank matrix, similar to that in \citet{Hamill-hybrid}; and (c) reducing the dimension $p$ of the observation in the update formula, for instance, by assimilating the observation in a serial way (see, for example,  \citealp{Whitaker-ensemble}), or by assimilating the observation in the framework of local EnKF (see, for example, \citealp{Bocquet2011-ensemble,Ott-local}). Once the problem of singularity is solved so that the smallest eigenvalue of $\mathbf{A}$ becomes positive, a (large enough) multiplicative inflation factor can be introduced to make sure that (\ref{eq:lambda_min_no_inf}) holds.

Inequality (\ref{eq:lambda_min_no_inf}) provides insights of what the constraints there may be in choosing the inflation factor. In what follows, we study the problem in a slightly more general setting. Concretely, we consider a family of mean update formulae in the form of
\begin{linenomath*}
\begin{subequations} \label{eq:mean_update_general}
\begin{align}
\label{sub_eq: mean_update}& \hat{\mathbf{x}}^a = \hat{\mathbf{x}}^b + \mathbf{G} \left(\mathbf{y}^o - \mathbf{H} \hat{\mathbf{x}}^b  \right) \, , \\
\label{sub_eq: gain_mtx} & \mathbf{G} = \alpha \, \hat{\mathbf{C}}^b \mathbf{H}^T \left( \delta \, \mathbf{H} \hat{\mathbf{C}}^b \, \mathbf{H}^T + \gamma \, \mathbf{R} \right)^{-1} \, ,
\end{align}
\end{subequations}
\end{linenomath*}
where $\alpha$, $\delta$ and $\gamma$ are some positive coefficients, and $\mathbf{G}$ is the gain matrix which in general differs from the Kalman gain $\mathbf{K}$ in Eq. (\ref{eq:mean_update}) with the presence of these three extra coefficients. Without loss of generality, though, one may let $\alpha=1$ (e.g., by moving $\alpha$ inside the parentheses) so that the gain matrix is simplified to
\begin{linenomath*}
\begin{equation} \label{eq:simplified_gain_mtx}
 \mathbf{G} = \hat{\mathbf{C}}^b \mathbf{H}^T \left( \delta \, \mathbf{H} \hat{\mathbf{C}}^b \mathbf{H}^T + \gamma \, \mathbf{R} \right)^{-1} \, , ~\text{with}~ \delta > 0 ~\text{and}~ \gamma > 0  .
\end{equation}
\end{linenomath*}
If $\delta =1$, then $\mathbf{G}$ resembles the Kalman gain in the EnKF, with $1/\gamma$ being analogous to the multiplicative covariance inflation factor as used in \citet{Anderson-Monte}. In our discussion below, we first derive some inflation constraints in the general case with $\delta > 0$, and then examine the more specific situation with $\delta = 1$. It is expected that one can also obtain constraints for other types of inflations in a similar way, but the results themselves may be case-dependent.

Using Eqs. (\ref{sub_eq: mean_update}) and (\ref{eq:simplified_gain_mtx}) as the update formulae and with some algebra, the weighted residual is given by
\begin{linenomath*}
\begin{equation} \label{eq:modified_weighted_residual_update}
(\mathbf{R}^{-1/2} \hat{\mathbf{r}}^a) = \left[ \mathbf{I}_p - \mathbf{A} \left(\delta \, \mathbf{A} + \gamma  \mathbf{I}_p \right)^{-1} \right] (\mathbf{R}^{-1/2} \hat{\mathbf{r}}^b ) \, ,
\end{equation}
\end{linenomath*}
where $\hat{\mathbf{r}}^a$, $\hat{\mathbf{r}}^b$ and $\mathbf{A}$ are defined as previously. Let
\begin{linenomath*}
\begin{equation} \label{eq:phi_def}
\begin{split}
\Phi & \equiv \mathbf{I}_p - \mathbf{A} \left(\delta \, \mathbf{A} + \gamma  \mathbf{I}_p \right)^{-1} \\
     & = \dfrac{\delta -1}{\delta} \, \mathbf{I}_p + \dfrac{\gamma}{\delta} \, \left(\delta \, \mathbf{A} + \gamma  \mathbf{I}_p \right)^{-1} \, ,
\end{split}
\end{equation}
\end{linenomath*}
then one has
\begin{linenomath*}
\begin{equation} \label{eq:modified_residual_normal_update}
\Vert \hat{\mathbf{r}}^a \Vert_{\mathbf{R}} = \Vert \mathbf{R}^{-1/2} \hat{\mathbf{r}}^a \Vert_2 = \Vert \Phi \, (\mathbf{R}^{-1/2} \hat{\mathbf{r}}^b) \Vert_2 \, .
\end{equation}
\end{linenomath*}
For our purpose, the following two matrix inequalities are useful. Firstly, given a matrix $\mathbf{M}$ and a vector $\mathbf{z}$ with suitable dimensions, one has 
\begin{linenomath*}
\begin{equation} \label{eq:mtx_ineq_1}
\Vert \mathbf{M} \, \mathbf{z} \Vert_2 \leq \Vert \mathbf{M} \Vert_2 \, \Vert \mathbf{z} \Vert_2 \, ,
\end{equation}
\end{linenomath*}
where $\Vert \mathbf{M} \Vert_2$, the induced 2-norm of $\mathbf{M}$, is the maximum of the absolute singular values of $\mathbf{M}$, or equivalently, $\Vert \mathbf{M} \Vert_2$ is equal to the square root of the largest eigenvalue of $\mathbf{M} \,  \mathbf{M}^T$ \citep[ch.~5]{Horn1990-matrix}. Secondly, if in addition $\mathbf{M}$ is non-singular, then (see, e.g., \citealp{Grcar2010-linear} and the references therein)
\begin{linenomath*}
\begin{equation} \label{eq:mtx_ineq_2}
\Vert \mathbf{M}^{-1} \Vert_2^{-1} \; \Vert \mathbf{z} \Vert_2  \leq \Vert \mathbf{M} \mathbf{z} \Vert_2 \, .
\end{equation}
\end{linenomath*}

The first inequality, (\ref{eq:mtx_ineq_1}), can be applied to obtain the sufficient conditions under which the inequality $\Vert \hat{\mathbf{r}}^a \Vert_{\mathbf{R}} \leq \beta_u \, \sqrt{p}$ is achieved. Let the maximum and minimum eigenvalues of $\Phi$ be $\mu_{max}$ and $\mu_{min}$, respectively. Then by Eq. (\ref{eq:phi_def})
\begin{linenomath*}
\begin{subequations} \label{eq:max_and_min_egen_of_phi}
\begin{align}
\label{subeq:max_egen_of_phi} & \mu_{max} = \dfrac{\delta -1}{\delta} + \dfrac{\gamma}{\delta} \, \left(\delta \, \lambda_{min} + \gamma \right)^{-1} \, , \\
\label{subeq:min_egen_of_phi} & \mu_{min} = \dfrac{\delta -1}{\delta} + \dfrac{\gamma}{\delta} \, \left(\delta \, \lambda_{max} + \gamma \right)^{-1} \, .
\end{align}
\end{subequations}
\end{linenomath*}
We remark that both $\mu_{max}$ and $\mu_{min}$ can be negative (e.g., when $\delta < 1$ and $\gamma \rightarrow 0$), therefore $\Vert \Phi \Vert_2 = \max(|\mu_{max}|,|\mu_{min}|) $. By (\ref{eq:modified_residual_normal_update}) and (\ref{eq:mtx_ineq_1}), a sufficient condition for $\Vert \hat{\mathbf{r}}^a \Vert_{\mathbf{R}} \leq \beta_u \, \sqrt{p}$ is $\max(|\mu_{max}|,|\mu_{min}|) \leq \beta_u \, \sqrt{p} / \Vert \hat{\mathbf{r}}^b \Vert_{\mathbf{R}}$. For notational convenience, we define $ \xi_u \equiv \beta_u \, \sqrt{p} / \Vert \hat{\mathbf{r}}^b \Vert_{\mathbf{R}}$ and $ \xi_l \equiv \beta_l \, \sqrt{p} / \Vert \hat{\mathbf{r}}^b \Vert_{\mathbf{R}}$.

Depending on the signs and magnitudes of $\mu_{max}$ and $\mu_{min}$, there are in general four possible scenarios: (a) $\mu_{max} \geq 0$ and $\mu_{min} \geq 0$, so that $\Vert \Phi \Vert_2 = \mu_{max}$; (b) $\mu_{max} \leq 0$ and $\mu_{min} \leq 0$, so that $\Vert \Phi \Vert_2 = - \mu_{min}$; (c) $\mu_{max} \geq 0$, $\mu_{min} \leq 0$ and $\mu_{max} + \mu_{min} \geq 0$, so that $\Vert \Phi \Vert_2 = \mu_{max}$; and (d) $\mu_{max} \geq 0$, $\mu_{min} \leq 0$ and $\mu_{max} + \mu_{min} \leq 0$, so that $\Vert \Phi \Vert_2 = - \mu_{min}$. Inserting Eq. (\ref{eq:max_and_min_egen_of_phi}) into the above conditions one obtains some inequalities with respect to the variables $\delta$ and $\gamma$ (subject to $\delta > 0$ and $\gamma > 0$), which are omitted in this note for brevity.

Similarly, the second inequality, (\ref{eq:mtx_ineq_2}), can be used to find the sufficient conditions for $ \beta_l \, \sqrt{p} \leq \Vert \hat{\mathbf{r}}^a \Vert_{\mathbf{R}} $. By (\ref{eq:modified_residual_normal_update}) and (\ref{eq:mtx_ineq_2}), one such sufficient condition can be $\Vert \Phi^{-1} \Vert_2 \leq  \Vert \hat{\mathbf{r}}^b \Vert_{\mathbf{R}} / (\beta_l \, \sqrt{p}) = 1/\xi_l $. By Eq. (\ref{eq:phi_def}) it can be shown that
\begin{linenomath*}
\begin{equation} \label{eq:inv_phi_def}
\Phi^{-1} =  \mathbf{I}_p + \left( (\delta -1) \, \mathbf{I}_p + \gamma \, \mathbf{A}^{-1} \right)^{-1} \, .
\end{equation}
\end{linenomath*}
Let the maximum and minimum eigenvalues of $\Phi^{-1} $ be $\nu_{max}$ and $\nu_{min}$, respectively, then
\begin{linenomath*}
\begin{subequations} \label{eq:max_and_min_egen_of_inv_phi}
\begin{align}
\label{subeq:max_egen_of_inv_phi} & \nu_{max} = 1 + \lambda_{max} \, \left( (\delta -1) \, \lambda_{max} + \gamma \right)^{-1} \, , \\
\label{subeq:min_egen_of_inv_phi} & \nu_{min} = 1 + \lambda_{min} \, \left( (\delta -1) \, \lambda_{min} + \gamma \right)^{-1} \, .
\end{align}
\end{subequations}
\end{linenomath*}
Similar to the previous discussion, we require that $\Vert \Phi^{-1} \Vert_2 = \max(|\nu_{max}|,|\nu_{min}|) \leq  1/\xi_l$, which also leads to four possible scenarios: (a) $\nu_{max} \geq 0$ and $\nu_{min} \geq 0$, so that $\Vert \Phi^{-1} \Vert_2 = \nu_{max}$; (b) $\nu_{max} \leq 0$ and $\nu_{min} \leq 0$, so that $\Vert \Phi^{-1} \Vert_2 = - \nu_{min}$; (c) $\nu_{max} \geq 0$, $\nu_{min} \leq 0$ and $\nu_{max} + \nu_{min} \geq 0$, so that $\Vert \Phi^{-1} \Vert_2 = \nu_{max}$; and (d) $\nu_{max} \geq 0$, $\nu_{min} \leq 0$ and $\nu_{max} + \nu_{min} \leq 0$, so that $\Vert \Phi^{-1} \Vert_2 = - \nu_{min}$. Again, inserting Eq. (\ref{eq:max_and_min_egen_of_inv_phi}) into the above conditions one obtains some inequalities with respect to the variables $\delta$ and $\gamma$.

Despite the complexity in the general situation, the analysis in the case of $\delta = 1$ (corresponding to the update formula in the EnKF) is significantly simplified. Indeed, when $\delta = 1$, the maximum and minimum eigenvalues in Eqs. (\ref{eq:max_and_min_egen_of_phi}) and (\ref{eq:max_and_min_egen_of_inv_phi}) are all positive. Therefore the following conditions
\begin{linenomath*}
\begin{subequations} \label{eq:enkf_sufficient_conditions}
\begin{align}
\label{subeq:sc_upper}& \mu_{max} = \gamma \, \left( \lambda_{min} + \gamma \right)^{-1} \leq \xi_u \, , \\
\label{subeq:sc_lower}& \nu_{max} = 1 + \lambda_{max} / \gamma \leq 1/ \xi_l  \, .
\end{align}
\end{subequations}
\end{linenomath*}
are sufficient for the objective $\beta_l \, \sqrt{p} \leq \Vert \hat{\mathbf{r}}^a \Vert_{\mathbf{R}} \leq \beta_u \, \sqrt{p}$. Note that if $\xi_u \geq 1$, i.e., $\Vert \hat{\mathbf{r}}^b \Vert_{\mathbf{R}} \leq \beta_u \, \sqrt{p}$, then any $\gamma > 0$ would guarantee that $\Vert \hat{\mathbf{r}}^a \Vert_{\mathbf{R}} \leq \beta_u \, \sqrt{p}$ (indeed by Eqs. (\ref{eq:modified_weighted_residual_update}) and (\ref{eq:mtx_ineq_1}) the analysis residual norm $\Vert \hat{\mathbf{r}}^a \Vert_{\mathbf{R}}$ is guaranteed to be no larger than $\Vert \hat{\mathbf{r}}^b \Vert_{\mathbf{R}}$ since $\Vert \Phi \Vert_2 \leq 1$ with $\delta = 1$), and that inequality (\ref{subeq:sc_upper}) holds. On the other hand, if $\xi_l \geq 1$ such that $\Vert \hat{\mathbf{r}}^b \Vert_{\mathbf{R}} \leq \beta_l \, \sqrt{p}$, then in most cases\footnote{An exception is in the case that $\gamma = + \infty$ and $\xi_l = 1$. This implies that $\Vert \hat{\mathbf{r}}^a \Vert_{\mathbf{R}} = \Vert \hat{\mathbf{r}}^b \Vert_{\mathbf{R}} = \beta_l \, \sqrt{p}$, and that no mean update is conducted (i.e., $\hat{\mathbf{x}}^a = \hat{\mathbf{x}}^b$).} it is impossible for the EnKF to have $\Vert \hat{\mathbf{r}}^a \Vert_{\mathbf{R}}$ no less than $\Vert \hat{\mathbf{r}}^b \Vert_{\mathbf{R}}$ (hence $\beta_l \, \sqrt{p}$), for the same aforementioned reason. Therefore the inequality (\ref{subeq:sc_lower}) becomes infeasible. With these said, in what follows we focus on the cases in which $\xi_u, \xi_l \in [0,1)$. With some algebra, it can be shown that $\gamma$ should be bounded by
\begin{linenomath*}
\begin{equation} \label{eq:constraints_on_gamma_enkf}
\dfrac{\xi_l}{1 - \xi_l} \, \lambda_{max}  \leq \gamma \leq \dfrac{\xi_u}{1 - \xi_u} \, \lambda_{min} \, .
\end{equation}
\end{linenomath*}
Let $\kappa = \lambda_{max} / \lambda_{min}$ be the condition number of the (normalized) matrix $\mathbf{A} = \mathbf{R}^{-1/2}\mathbf{H} \hat{\mathbf{C}}^b \mathbf{H}^T \mathbf{R}^{-T/2}$. From (\ref{eq:constraints_on_gamma_enkf}) we have $\dfrac{\xi_l}{1 - \xi_l} \, \lambda_{max} \leq \dfrac{\xi_u}{1 - \xi_u} \, \lambda_{min}$, which leads to a constraint in choosing $\beta_l$ and $\beta_u$, in terms of
\begin{linenomath*}
\begin{equation} \label{eq:constraints_on_beta_l}
\beta_l \leq \dfrac{\beta_u}{\kappa + (1-\kappa) \, \xi_u} \, .
\end{equation}
\end{linenomath*}

Inequality (\ref{eq:constraints_on_gamma_enkf}) suggests that the upper and lower bounds of $\gamma$ are related to the minimum and maximum eigenvalues of $\mathbf{A}$, respectively. In particular, to avoid a too small residual norm, i.e., observation over-fitting, $\gamma$ should be lower bounded, hence its inverse $1/\gamma$, resembling the multiplicative inflation factor, should be upper bounded, as mentioned previously. 

In practice, if the dimension $p$ of the observation space is large, then it may be expensive to evaluate $\lambda_{max}$ and $\lambda_{min}$. In certain circumstances, though, there may be cheaper ways to compute an interval for $\gamma$. For instance, if $\hat{\mathbf{C}}^b$ in the mean update formula is in the form of $c_1 \, \hat{\mathbf{P}}^b + c_2 \, \mathbf{B}$ with $c_1$ and $c_2$ being some positive scalars and $\mathbf{B}$ a constant, symmetric and positive-definite matrix, then
\begin{linenomath*}
\[
\mathbf{A} = c_1 \, \mathbf{R}^{-1/2}\mathbf{H} \hat{\mathbf{P}}^b \mathbf{H}^T \mathbf{R}^{-T/2} + c_2 \, \mathbf{R}^{-1/2}\mathbf{H} \mathbf{B} \mathbf{H}^T \mathbf{R}^{-T/2} \, .
\]
\end{linenomath*}
The additive Weyl inequality \citep[ch.~3]{horn1991topics} suggests that the following bounds hold for $\lambda_{max}$ and $\lambda_{min}$.
\begin{linenomath*}
\begin{equation} \label{eq:bounds_on_lambda}
\begin{split}
& \lambda_{max} \leq c_1 \, \tau_{max} + c_2 \, \rho_{max} \, , \\
& \lambda_{min} \geq c_1 \, \tau_{min} + c_2 \, \rho_{min} \geq c_2 \, \rho_{min} \, , \\
\end{split}
\end{equation}
\end{linenomath*}
where $\tau$ and $\rho$ are the eigenvalues of $\mathbf{R}^{-1/2}\mathbf{H} \hat{\mathbf{P}}^b \mathbf{H}^T \mathbf{R}^{-T/2}$ and $\mathbf{R}^{-1/2}\mathbf{H} \mathbf{B} \mathbf{H}^T \mathbf{R}^{-T/2}$, respectively. In many situations, $\hat{\mathbf{P}}^b$ may be rank deficient, therefore a singular value decomposition (SVD) analysis shows that  $\tau_{max}$ is equal to the largest eigenvalue of $ (\mathbf{H} \hat{\mathbf{S}}^b)^T \mathbf{R}^{-1} (\mathbf{H} \hat{\mathbf{S}}^b)$, where $\hat{\mathbf{S}}^b$ is a square root of $\hat{\mathbf{P}}^b$ that can be directly constructed based on the background ensemble \citep{Bishop-adaptive,Luo-ensemble,Wang-which}. Note that $ (\mathbf{H} \hat{\mathbf{S}}^b)^T \mathbf{R}^{-1} (\mathbf{H} \hat{\mathbf{S}}^b)$ is a matrix with its dimension determined by the ensemble size $n$, and is in fact the same as the one used in the ensemble transform Kalman filter (ETKF) \citep{Bishop-adaptive,Wang-which} in order to obtain the transform matrix. Therefore $\tau_{max}$ can be taken as a by-product in the framework of ETKF. On the other hand, if both $\mathbf{H}$ and $\mathbf{R}$ are time-invariant, then the eigenvalues $\rho_{max}$ and $\rho_{min}$ of $\mathbf{R}^{-1/2}\mathbf{H} \mathbf{B} \mathbf{H}^T \mathbf{R}^{-T/2}$ can be calculated off-line once and for all. Taking these considerations into account, (\ref{eq:constraints_on_gamma_enkf}) can be modified as follows
\begin{linenomath*}
\begin{equation} \label{eq:modified_constraints_on_gamma_enkf}
\dfrac{\xi_l}{1 - \xi_l} \, (c_1 \, \tau_{max} + c_2 \, \rho_{max})  \leq \gamma \leq \dfrac{\xi_u}{1 - \xi_u} \, (c_2 \, \rho_{min}) \, .
\end{equation}
\end{linenomath*}
Accordingly, (\ref{eq:constraints_on_beta_l}) is changed to
\begin{linenomath*}
\begin{equation} \label{eq:modified_constraints_on_beta_l}
\beta_l \leq \dfrac{\beta_u}{\tilde{\kappa} + (1-\tilde{\kappa}) \, \xi_u} \, ,
\end{equation}
\end{linenomath*}
with $\tilde{\kappa} = (c_1 \, \tau_{max} + c_2 \, \rho_{max}) / (c_2 \, \rho_{min})$ being a modified ``condition number''.

\noindent \textbf{Remark:} Inequalities (\ref{eq:constraints_on_gamma_enkf}) and (\ref{eq:constraints_on_beta_l}), or alternatively, (\ref{eq:modified_constraints_on_gamma_enkf}) and (\ref{eq:modified_constraints_on_beta_l}), are sufficient, but not necessary, conditions. Therefore, even though $\gamma$ does not lie in the interval in (\ref{eq:constraints_on_gamma_enkf}) or (\ref{eq:modified_constraints_on_gamma_enkf}), it may be still possible for the analysis residual norm to satisfy $\beta_l \, \sqrt{p} \leq \Vert \hat{\mathbf{r}}^a \Vert_{\mathbf{R}} \leq \beta_u \, \sqrt{p}$.
\section{Numerical verification}

Here we focus on using the 40-dimensional Lorenz 96 (L96) model \citep{Lorenz-optimal} to verify the above analytic results, while more intensive filter (with residual nudging) performance investigations are reported in \citet{Luo2012-residual}. The experiment settings are the following. A reference trajectory (truth) is generated by numerically integrating the L96 model {\color{black}{(with the driving force term $F=8$)}} forward through the {\color{black}{fourth-order}} Runge-Kutta method, with the integration step being 0.05 and the total number of integration steps being 1500. The first 500 steps are discarded to avoid the transition effect, and the rest 1000 steps are used for data assimilation. To obtain a long-term ``background covariance'' ${\color{black}{\mathbf{B}^{lt}}}$ (``background mean'' $\mathbf{x}^{B}$, respectively), we also conduct a separate long model run with $100,000$ integration steps, and take ${\color{black}{\mathbf{B}^{lt}}}$ ($\mathbf{x}^{B}$) as the temporal covariance (mean) of the generated model trajectory. The synthetic observations are generated by adding the Gaussian white noise $N(0,1)$ to each odd number elements ($x_1,x_3,\dotsb,x_{39}$) of the state vector $\mathbf{x} = [x_1,x_2,\dotsb,x_{40}]^T$ every 4 integration steps. This corresponds to the $1/2$ observation scenario used in \citet{Luo2012-residual}. An initial ensemble with $20$ ensemble members is generated by drawing samples from the Gaussian distribution $N(\mathbf{x}^{B},{\color{black}{\mathbf{B}^{lt}}})$, and the ETKF is adopted for data assimilation.

For distinction later, we call the ETKF without residual nudging the normal ETKF, and the ETKF with residual nudging the ETKF-RN. In the normal ETKF, Eq. (\ref{eq:mean_update}) is used for mean update with $\hat{\mathbf{C}}^b$ equal to the sample error covariance $\hat{\mathbf{P}}^b$ of the background ensemble\footnote{\color{black}{One may also let $\hat{\mathbf{C}}^b$ be the hybrid of $\hat{\mathbf{P}}^b$ and $\mathbf{B}^{lt}$. In this case, both residual norms and root mean square errors (RMSEs) of the normal ETKF may become smaller (results not shown), while the validity of the analytic results in the previous section is not affected.}}. Neither covariance inflation nor covariance localization is introduced to the normal ETKF, since for our purpose we wish to use this plain filter setting as the baseline for comparison. One may adopt various inflation and localization techniques to enhance the filter performance, but such an investigation is beyond the scope of this note. 

In the ETKF-RN, we adopt the hybrid scheme $\hat{\mathbf{C}}^b = 0.5 \hat{\mathbf{P}}^b + 0.5 {\color{black}{\mathbf{B}^{lt}}}$ to address the issue of possible singularity in the matrix $\mathbf{A}$ (cf. Eq. \ref{eq:mtx_A}). Eq. (\ref{eq:mean_update_general}) is adopted for mean update, with $\alpha = \delta = 1$, and $\gamma$ constrained by (\ref{eq:modified_constraints_on_gamma_enkf}) and (\ref{eq:modified_constraints_on_beta_l}). For convenience, we denote the lower and upper bounds of $\gamma$ in (\ref{eq:modified_constraints_on_gamma_enkf}) by $\gamma_{min}$ and $\gamma_{max}$, respectively, and re-write $\gamma$ in terms of $\gamma = \gamma_{min} + c \, (\gamma_{max} - \gamma_{min})$ with $c$ being a corresponding scalar coefficient that is involved in our discussion later. {\color{black}{Note that in general the background residual norm $\Vert \hat{\mathbf{r}}^b \Vert_{\mathbf{R}}$ changes with time, so are the values of $\xi_u$ and $\xi_l$ in Eq. (\ref{eq:constraints_on_gamma_enkf}). This implies that in general $\gamma_{min}$ and $\gamma_{max}$ (hence $\gamma$) also change with time, therefore they need to be calculated at each data assimilation cycle.}}  

An additional remark is that the normal ETKF and the ETKF-RN share the same square root update formula as in \citet{Wang-which}, where it is the sample error covariance $\hat{\mathbf{P}}^b$, rather than its hybrid with ${\color{black}{\mathbf{B}^{lt}}}$, which is used to generate the background square root. {\color{black}{Such a choice is based on the following considerations. On the one hand, if one uses the hybrid covariance for square root update, then it would require a matrix factorization (e.g., singular value decomposition) in order to compute a square root of the hybrid covariance at each data assimilation cycle, which can be very expensive in large-scale applications. On the other hand, for the L96 model used here, numerical investigations show that using the hybrid covariance for square root update does not necessarily improve the filter performance (results not shown).}}

{\textcolor{black}{The procedures in the ETKF-RN are summarized as follows. Because the matrix $\mathbf{R}^{-1/2}\mathbf{H} \mathbf{B} \mathbf{H}^T \mathbf{R}^{-T/2}$ is time invariant, its maximum and minimum eigenvalues, $\rho_{max}$ and $\rho_{min}$ (cf. (\ref{eq:modified_constraints_on_gamma_enkf})), respectively, are calculated and saved for later use. Then, with the background ensemble at each data assimilation cycle, calculate the sample mean $\hat{\mathbf{x}}^b$, the corresponding background residual norm $\Vert \hat{\mathbf{r}}^b \Vert_{\mathbf{R}}$, and a square root $\hat{\mathbf{S}}^b$ of the sample error covariance $\hat{\mathbf{P}}^b$ following \citet{Bishop-adaptive,Luo-ensemble,Wang-which}. Update $\hat{\mathbf{S}}^b$ to its analysis counterpart $\hat{\mathbf{S}}^a \equiv \hat{\mathbf{S}}^b \mathbf{T} \mathbf{U}$ by calculating a transform matrix $\mathbf{T}$, together with a ``centering'' matrix $\mathbf{U}$ following \citet{Wang-which}. During the square root update process, the maximum eigenvalue $\tau_{max}$ of $\mathbf{R}^{-1/2}\mathbf{H} \hat{\mathbf{P}}^b \mathbf{H}^T \mathbf{R}^{-T/2}$ is obtained as a by-product following our discussion in the previous section. With these information, one is ready to calculate the interval bounds $\gamma_{min}$ and $\gamma_{max}$ in (\ref{eq:modified_constraints_on_gamma_enkf}), hence obtain $\gamma = \gamma_{min} + c \, (\gamma_{max} - \gamma_{min})$ for a given value of $c$ ($c$ can be constant or variable during the whole data assimilation time window). This $\gamma$ value is then inserted into Eq. (\ref{eq:mean_update_general}) (with $\alpha = \delta = 1$ there) to obtain the analysis mean $\hat{\mathbf{x}}^a$. With $\hat{\mathbf{x}}^a$ and $\hat{\mathbf{S}}^a$, an analysis ensemble can be generated in the same way as in \citet{Bishop-adaptive,Wang-which}. Propagating this ensemble forward in time, one starts a new data assimilation cycle, and so on. Comparing the above procedures to those in \citet{Luo2012-residual}, the observation inversion used in \citet{Luo2012-residual} is avoided.}} 

The experiment below aims to show that, at each data assimilation cycle, if a $\gamma$ value lies in the interval $\mathbb{C}_{\gamma} = [\gamma_{min}, \gamma_{max}]$ given by (\ref{eq:modified_constraints_on_gamma_enkf}), then the corresponding analysis residual norm $\Vert \hat{\mathbf{r}}^a \Vert_{\mathbf{R}}$ is bounded by the interval $\mathbb{C}_{rn} = [\beta_l \sqrt{p}, \beta_u  \sqrt{p}]$, with $\beta_l$ and $\beta_u$ satisfying the constraint (\ref{eq:modified_constraints_on_beta_l}). In the experiment we fix $\beta_u = 2$, and let $\beta_l = 0.1 \times (\beta_u / (\tilde{\kappa} + (1-\tilde{\kappa}) \, \xi_u))$, where the small fraction $0.1$ is introduced for convenience of visualization\footnote{In some cases $\beta_u / (\tilde{\kappa} + (1-\tilde{\kappa}) \, \xi_u)$ in (\ref{eq:modified_constraints_on_beta_l}) may be very close to $\beta_u$. Therefore if $\beta_l$ is close to this value, the difference $(\beta_u - \beta_l)$, hence the interval $\mathbb{C}_{rn}$, may be very small.}.

\begin{figure*} 
\vspace*{2mm}
\centering

\subfigure[]{ \label{subfig:normal_output}
\includegraphics[scale = 0.43]{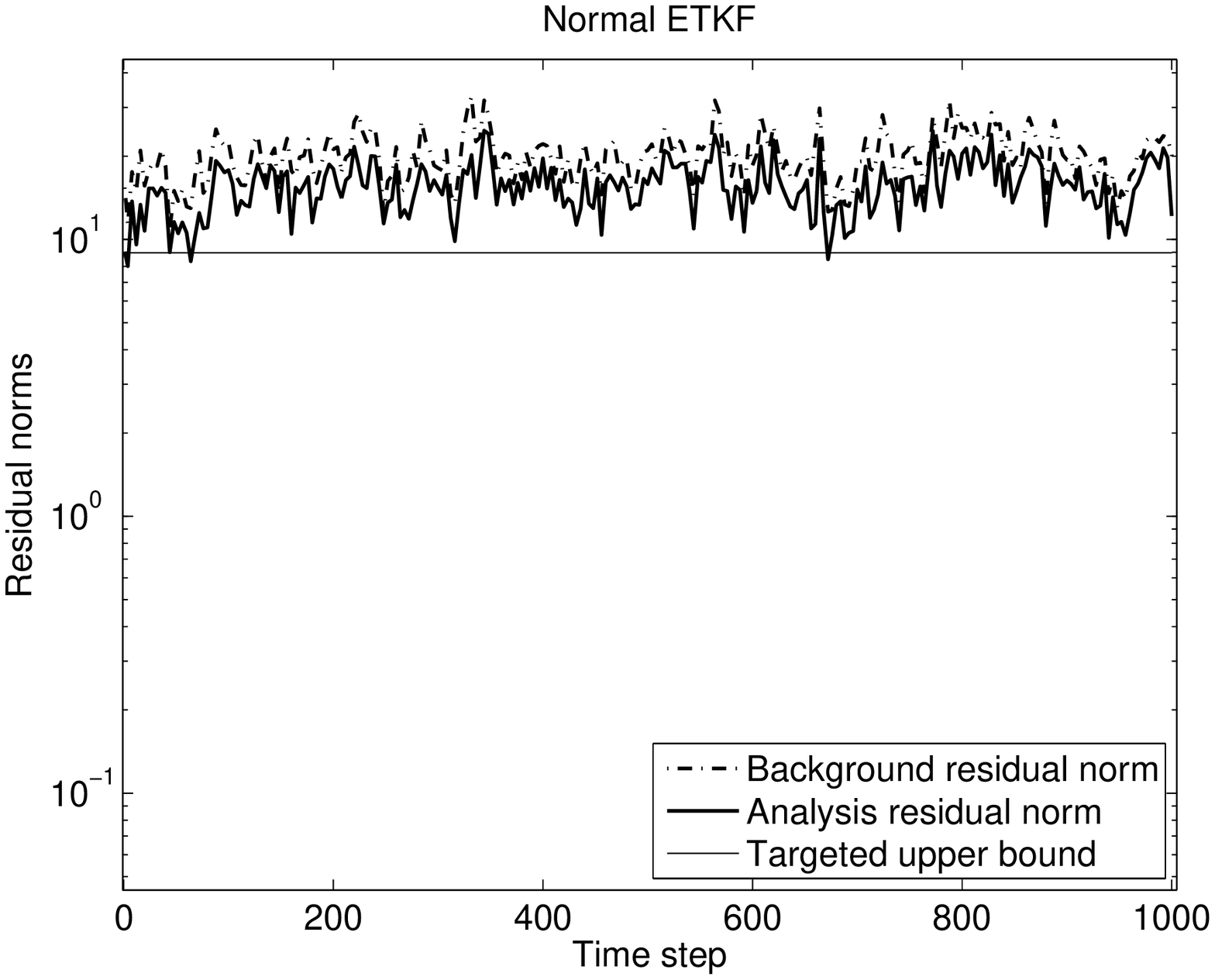}
}
\subfigure[]{ \label{subfig:darn_output_0}
\includegraphics[scale = 0.43]{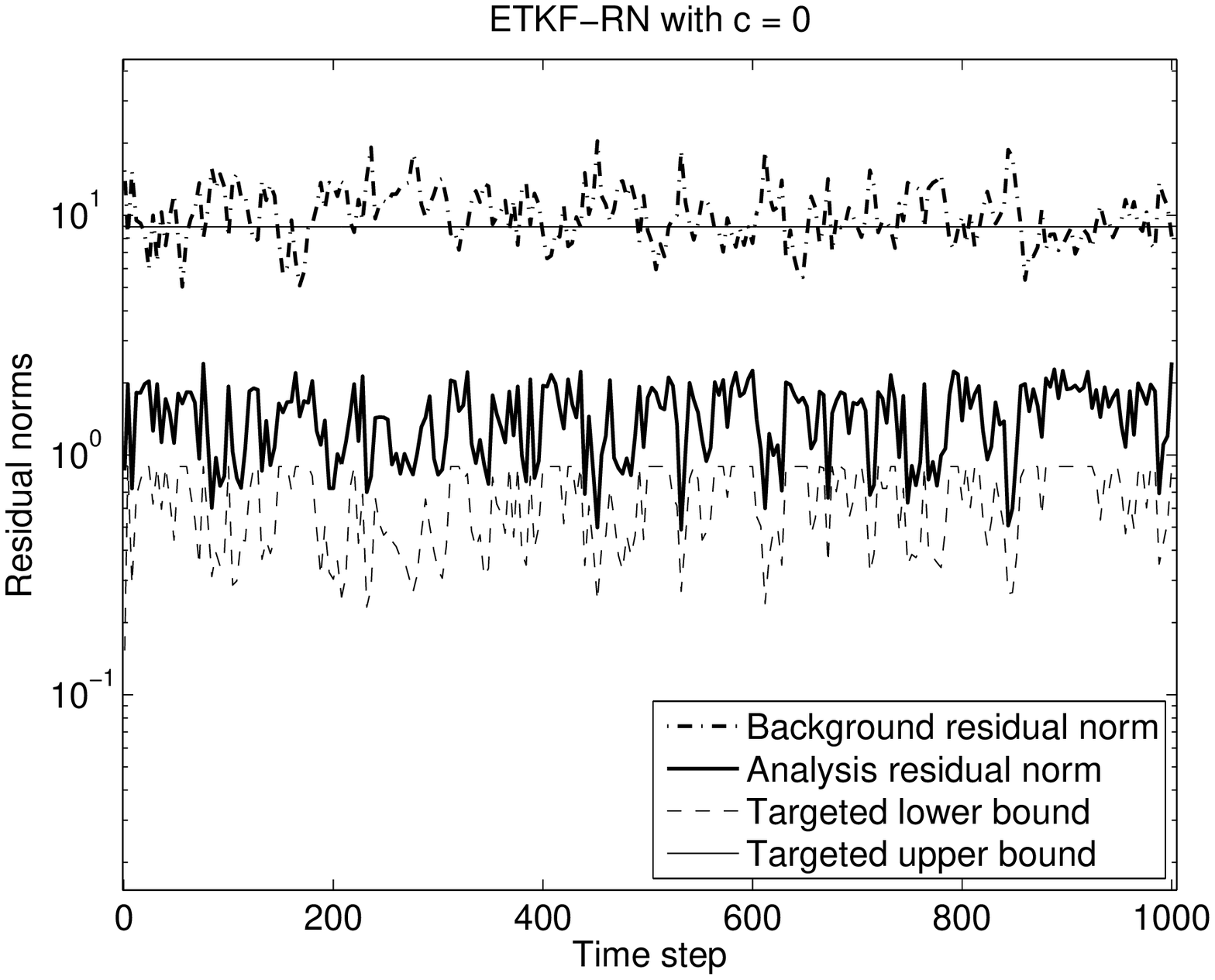}
}
\subfigure[]{ \label{subfig:darn_output_1}
\includegraphics[scale = 0.43]{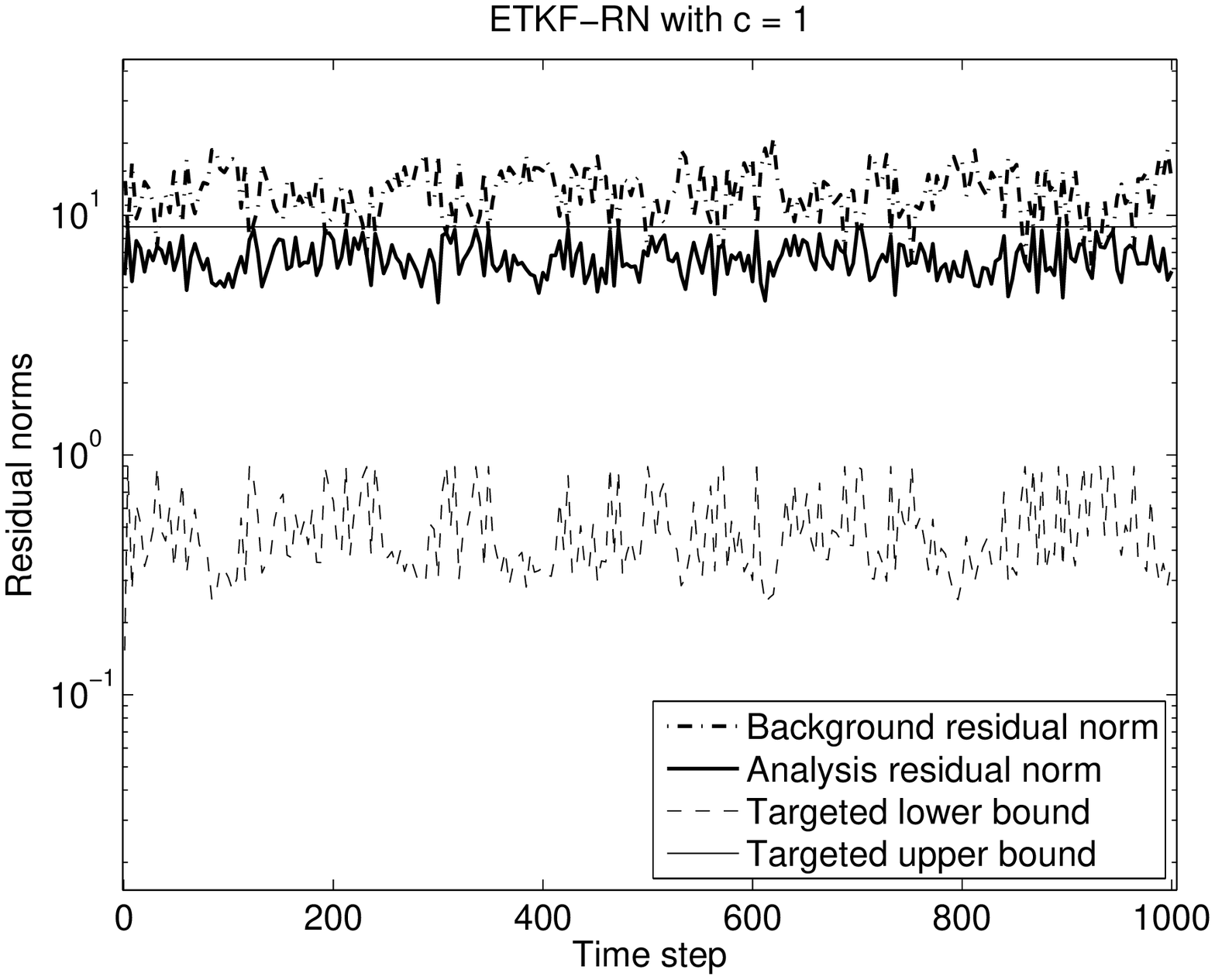}
}
\subfigure[]{ \label{subfig:darn_output_random}
\includegraphics[scale = 0.43]{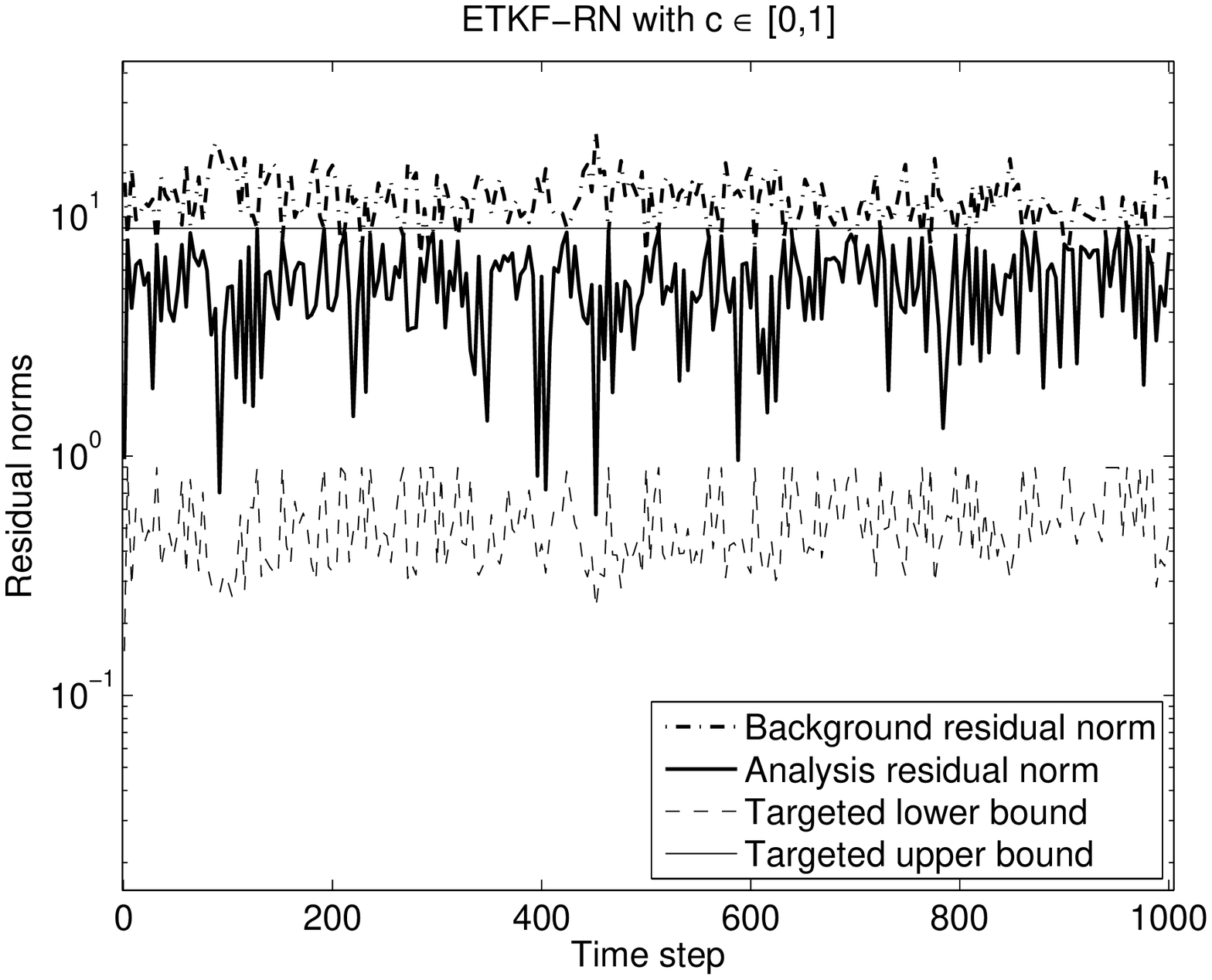}
}
\subfigure[]{ \label{subfig:darn_output_2.5}
\includegraphics[scale = 0.43]{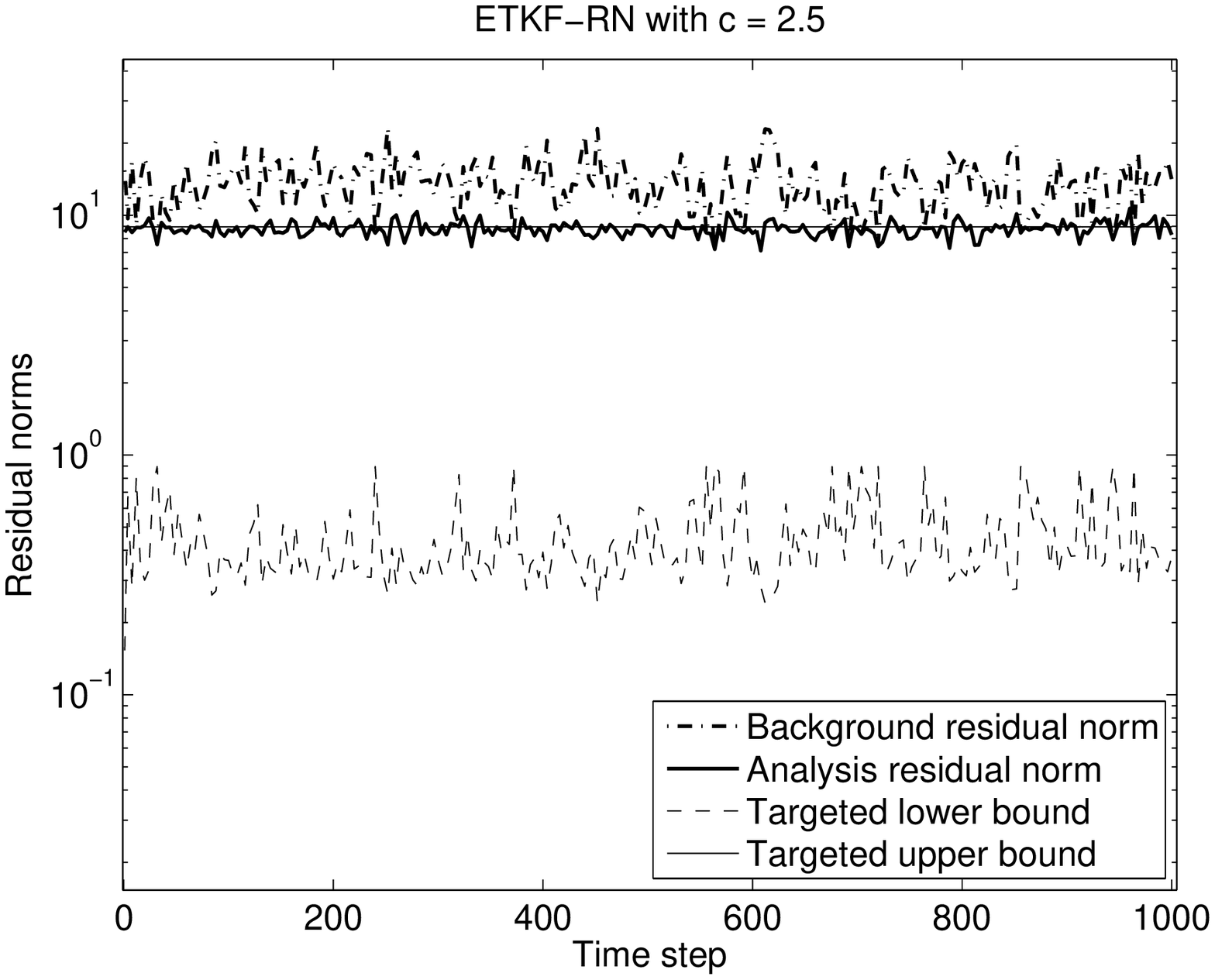}
}
\subfigure[]{ \label{subfig:darn_output_-0.005}
\includegraphics[scale = 0.43]{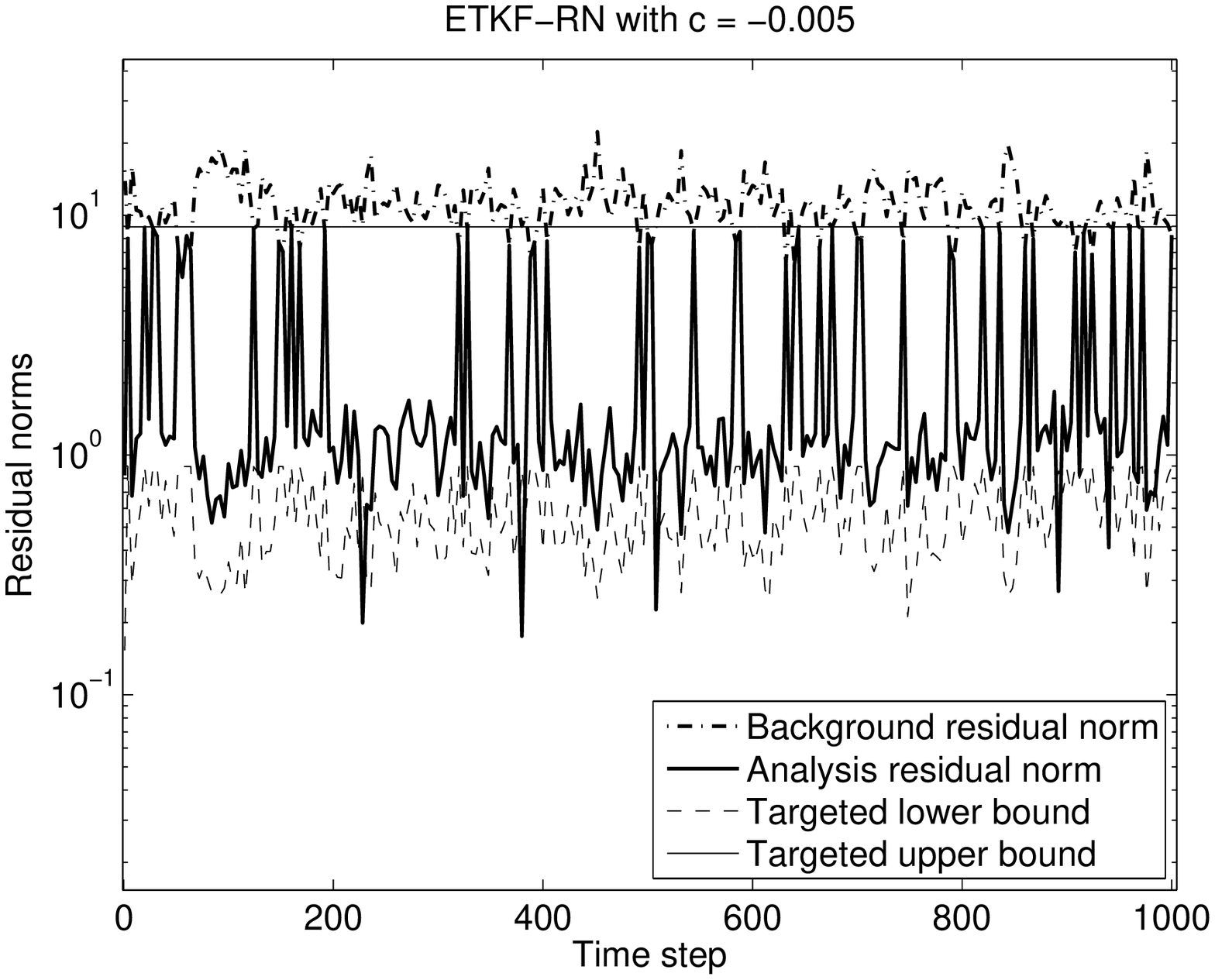}
}

\caption{\label{fig:output_residual_norm} Time series of the analysis residual norms in: (a): the normal ETKF without residual nudging; (b) -- (f) the ETKF-RN with different $c$ values. For the normal ETKF there are no targeted lower and upper residual norm bounds. For reference, though, we still plot the targeted upper bound ($=2\sqrt{20}$) in (a). {\color{black}{We also note that the $c$ value in Fig. \ref{subfig:darn_output_random} is randomly drawn from the uniform distribution on the interval $[0,1]$ at each data assimilation cycle, while in the rest of the sub-figures the $c$ values are constant during the assimilation time window.}}}
\end{figure*}

Fig. \ref{fig:output_residual_norm} shows the time series of the background (dash-dotted) and analysis (thick solid) residual norms in different filter settings (for convenience of visualization, the residual norm values are plotted in the logarithmic scale). For reference we also plot the targeted lower and upper bounds (dash and thin solid lines, respectively), $\beta_l \sqrt{p}$ and  $\beta_u \sqrt{p}$ ($p=20$), respectively. In the normal ETKF (Fig. \ref{subfig:normal_output}), in most of the time the analysis residual norms are larger than the targeted upper bound (no targeted lower bound is calculated and plotted in this case). With residual nudging, the analysis residual norms of the ETKF-RN migrate into the targeted interval, as long as the coefficient $c$ lies in $[0,1]$ (Figs. \ref{subfig:darn_output_0} -- \ref{subfig:darn_output_random}. {\color{black}{Also see the caption of Fig. \ref{fig:output_residual_norm} to find out how the corresponding $c$ values are chosen}}). When $c$ is outside the interval $[0,1]$, the corresponding $\gamma$ is not bounded by $[\gamma_{min},\gamma_{max}]$, hence there is no guarantee that the corresponding analysis residual norms are bounded by $[\beta_l \sqrt{p},\beta_u \sqrt{p}]$. Two such examples are presented in Fig. \ref{subfig:darn_output_2.5} and \ref{subfig:darn_output_-0.005}, with $c$ being $2.5$ and $-0.005$, respectively (e.g., for $c = -0.005$ in Fig. \ref{subfig:darn_output_-0.005}, breakthroughs of the lower bound are found around time step $220$ and a few other places). {\color{black}{As side results, we also report in Table \ref{table:rmse} the time mean root mean square errors (RMSEs) (see Eq. (13) of \citealp{Luo2012-residual}) that correspond to different filter settings in Fig. \ref{fig:output_residual_norm}.}} In these tested cases, the filter performance of the ETKF-RN appears improved, in terms of the time mean RMSE, when compared to that of the normal ETKF. 
\section{Discussion and conclusion}

We derived some sufficient inflation constraints in order for the analysis residual norm to be bounded in a certain interval. The analytic results showed that these constraints are related to the maximum and minimum eigenvalues of certain matrices (cf. Eq. (\ref{eq:mtx_A})). In certain circumstances, the constraint with respect to the minimum eigenvalue (e.g., Eq. (\ref{eq:lambda_min_no_inf})) may impose a non-singularity requirement on relevant matrices. A few strategies in the literature that can be adopted to address or mitigate this issue are highlighted.

Some remaining issues are manifest in our deduction. These include, for instance, the nonlinearity in the observation operator and the choice of $\beta_u$ and $\beta_l$. For the former problem, under a suitable smoothness assumption on the observation operator, one may also obtain inflation constraints similar to those in Section \ref{sec:inflation}. On the other hand, though, more investigations may be needed to make the results more practical in terms of computational complexity. For the latter problem,   numerical results in \citet{Luo2012-residual} show that the $\beta$ values influence the overall performance of the EnKF in terms of filter stability and accuracy. Intuitively, smaller (larger) $\beta$ values tend to make residual nudging happen more (less) often. Therefore, if the normal EnKF performs well (poorly), then a larger (smaller) $\beta$ value may be suitable. In this aspect, it is expected that an objective criterion is needed. This will be investigated in the future.

\section*{Acknowledgement}

We would like to thank two anonymous reviewers for their constructive comments and suggestions. The first author would also like to thank the IRIS/CIPR cooperative research project ``Integrated Workflow and Realistic Geology'' which is funded by industry partners ConocoPhillips, Eni,  Petrobras, Statoil, and Total, as well as the Research Council of Norway (PETROMAKS) for financial support.

\bibliographystyle{ametsoc}
\bibliography{./references}

\clearpage
\begin{table*} 
\centering
\caption{\label{table:rmse} Time mean RMSEs in the normal ETKF and the ETKF-RN with the same c values as in Fig. \ref{fig:output_residual_norm}.}
\begin{tabular}{lllllll}
\hline \hline
& \multirow{2}{*}{Normal ETKF} & \multicolumn{5}{c}{ETKF-RN with} \\
\cline{3-7}
& & $c=0$ & $c=1$ & $c \in [0,1]$ & $c=2.5$ & $c=-0.005$ \\
\hline
Background RMSE & 4.3148 & 1.8252 & 2.4095 & 2.2182 & 2.6857 & 2.0394 \\
Analysis RMSE & 4.2645 & 1.6953 & 2.2764 & 2.0894 & 2.5679 & 1.9054 \\
\hline \hline
\end{tabular}
\end{table*}

\clearpage




\end{document}